\documentclass[%
 reprint,
 amsmath,amssymb,
 aps,
]{revtex4-1}

\usepackage[brazilian, english]{babel}
\usepackage[svgnames]{xcolor} % Specify colors by their 'svgnames', for a full list of all colors available see here: http://www.latextemplates.com/svgnames-colors
\usepackage{graphicx}% Include figure files
\graphicspath{{figures/}} % Location of the graphics files
\usepackage[font=small,labelfont=bf, justification=raggedright,singlelinecheck=false]{caption} % Required for specifying captions to tables and figures
\usepackage{amsfonts, amsmath, amsthm, amssymb} % For math fonts, symbols and environments
\usepackage{wrapfig} % Allows wrapping text around tables and figures
\usepackage{dcolumn}% Align table columns on decimal point
\usepackage{bm}% bold math
\usepackage{footnote}
\usepackage{tabularx}
\usepackage{float}
\usepackage{braket}
\usepackage[utf8]{inputenc}
\begin{document}

%\preprint{APS/123-QED}

\title{Layer breathing and shear modes in multilayer graphene: A DFT-vdW study}% Force line breaks with \\

\author{Rafael R. Del Grande}
\email{rdgrande@if.ufrj.br}
\affiliation{%
Instituto de Física, Universidade Federal do Rio de Janeiro, 
Caixa Postal 68528, Rio de Janeiro, RJ 21941-972, Brazil}%
\author{Marcos G. Menezes}
\email{marcosgm@if.ufrj.br}
\affiliation{%
Instituto de Física, Universidade Federal do Rio de Janeiro, 
Caixa Postal 68528, Rio de Janeiro, RJ 21941-972, Brazil}%
\author{Rodrigo B. Capaz}
\email{capaz@if.ufrj.br}
\affiliation{%
Instituto de Física, Universidade Federal do Rio de Janeiro, 
Caixa Postal 68528, Rio de Janeiro, RJ 21941-972, Brazil}%

\date{\today}% It is always \today, today,

\begin{abstract}

In this work, we study structural and vibrational properties of multilayer graphene using density-functional theory (DFT) with van der Waals (vdW) functionals. Initially, we analyze how different vdW functionals compare by evaluating the lattice parameters, elastic constants and vibrational frequencies of low energy optical modes of graphite. Our results indicate that the vdW-DF1-optB88 functional has the best overall performance on the description of vibrational properties. Next, we use this functional to study the influence of the vdW interactions on the structural and vibrational properties of multilayer graphene. Specifically, we evaluate binding energies, interlayer distances and phonon frequencies of layer breathing and shear modes. We observe excellent agreement between our calculated results and available experimental data, which suggests that this functional has truly predictive power for layer-breathing and shear frequencies that have not been measured yet. This indicates that careful selected vdW functionals can describe interlayer bonding in graphene-related systems with good accuracy.

\end{abstract}

\maketitle

%\tableofcontents

\section{\label{sec:level1}Introduction}

Since the experimental realization of graphene \cite{novoselov2004Science, Novoselov2005Nat, geim2007NatMat, GeimScience2009, Novoselov2011RMP}, an extensive number of studies have been made on this material and related structures due to their remarkable properties and potential applications \cite{Neto2009RevModPhys, Novoselov2016Science, Ferrari2015Nanoscale, Butler2013ACSnano, Geim2013Nat}. In particular, one class of graphene related materials that is of great interest is multilayer graphene. $N$-layer graphene (NLG) consists of $N$ stacked graphene layers, which can have different orientations with respect to each other. Their mechanical, electronic and optical properties strongly depend on $N$ and their relative \cite{MenezesPRB2014, VelaPRB2018} orientations, resulting in a wide variety of potential applications \cite{cao2018Nature, Munier2016RevModPhys}.

These materials are stable due to the van der Waals (vdW) interactions between layers (as in graphite). Therefore, in order to study them from a theoretical point of view, one needs a good description of these dispersive interactions, specially for properties related to interlayer bonding, stiffness and vibrations. In particular, the effects of vdW interactions can be detected in Raman spectroscopy, as interlayer vibrations in 2D materials can be measured by this technique with great precision. For this reason, Raman scattering may be used as a tool to evaluate the accuracy of various theoretical implementations of vdW interactions in density-functional theory (DFT) calculations \cite{EklundCarbon1995, TanNatMat2012, LuiNanoLet2014, Lui2013PRB, Wu2014NatComm, Boschetto2013NanoLet}. 
\\

In DFT, the exact exchange-correlation functional should in principle contain vdW interactions.
However, the exact functional is unknown, and commonly used local (LDA) and semilocal (GGA) approximations do not capture the effects of vdW interactions, which are dispersion forces arising from long-range electron-electron correlations \cite{berland2015RepProgPhys, langreth2005IntJourQChem}. In graphite, it is known that GGA functionals considerably overestimate the interlayer distance, while describing well the in-plane carbon-carbon distances. On the other hand, LDA functionals give a good interplanar distance (although smaller than the experimental value) and underestimates the covalent bond lengths \cite{MounetPRB2005}.

In recent works, non-local vdW exchange-correlation functionals were developed aiming to describe the properties of layered materials, adsorption processes and other phenomena \cite{Dion2005PRB, thonhauser2015PRB, lee2010PRB, klimes2010JPCondMatter, klimes2011PRB, cooper2010PRB, berland2014PRB, hamada2014PRB, Sabatini2013PRB}. %adicionar referências 
In particular, in the vdW-DF family of functionals \cite{Dion2005PRB}, the exchange-correlation energy is given by:

\begin{equation}
E_{xc}[n] = E_x^{GGA}[n] + E_c^{LDA}[n] + E_c^{nl}[n]
\end{equation}

\noindent
where \textit{x} and \textit{c} label the exchange and correlation parts of the corresponding functionals, respectively, and $E_c^{nl}[n]$ is a non-local correlation energy given by

\begin{equation}
E_c^{nl}[n] = \int d^3r d^3r' n(\bold{r}) \phi[n](\bold{r}, \bold{r'}) n(\bold{r'}) 
\label{Non_local_corr_energy}
\end{equation}

\noindent
where $\phi[n]$ is a two-point kernel function and $n(\bold{r})$ is the electronic density. 
The different functionals in this family consist of different combinations of flavors for the GGA exchange and LDA correlation, as well as different forms for the two-point kernel \cite{Dion2005PRB, thonhauser2015PRB, lee2010PRB, klimes2010JPCondMatter, klimes2011PRB, cooper2010PRB, berland2014PRB, hamada2014PRB, Sabatini2013PRB}. They were tailored to describe properties of specific systems, such as the interaction energy of dimer pairs and the structure of small molecules. Therefore, their performances can dramatically change depending on chemical environment. For a complete overview see Refs. \onlinecite{berland2015RepProgPhys} and \onlinecite{VdwDFBook}.

In this work, we compare the performances of different vdW functionals  from the DF family in describing the elastic constants, binding energy and vibrational frequencies of the shear and layer breathing modes (LBM) of graphite. Our results indicate that the vdW-DF1-optB88 functional is the best suited functional for graphene-based systems. Next, we use this functional to calculate interlayer distances, binding energies and phonon frequencies of NLG. We find that the inclusion of vdW interactions is especially important for an accurate description of interlayer distances and out-of-plane vibrational modes.

\section{\label{sec:level1}Methods}

All DFT calculations were performed using the Quantum Espresso package \cite{Giannozzi2009PhysCondMatter, GiannozziJourPhys2017}. Energy cutoffs of $60$ and $480$ Ry are used for the plane-wave expansion of the wavefunctions and electronic density, respectively. For graphite calculations, several functionals are used.
For LDA and GGA calculations, we use the PZ and PBE functionals, respectively \cite{perdew1996PRL, perdew1981PRB}. For vdW calculations, we employ different functionals from the vdW-DF family \cite{Dion2005PRB, thonhauser2015PRB, lee2010PRB, klimes2010JPCondMatter, klimes2011PRB, cooper2010PRB, berland2014PRB, hamada2014PRB, Sabatini2013PRB}.
In all cases, we use RRKJ ultrasoft pseudopotentials for the ion-electron interaction, including a non-linear core correction \cite{Rappe1990PRB}. It should be noted that, so far, pseudopotentials for non-local functionals have not been developed, so we use a GGA-derived potential for the vdW calculations as in \cite{hamada2014PRB}.

Structural optimizations are done with a convergence threshold on forces of $10^{-6}$ Ry/bohr and on energies of $10^{-5}$ Ry. To sample the Brillouin Zone, we have used Monkhorst-Pack meshes of dimensions $16 \times 16 \times 16$ for graphite and $16 \times 16 \times 1$ for NLG. 

The elastic constants of graphite are calculated from second order derivatives of the total energy with respect to the lattice parameters \cite{BoettgerPRB1997, MounetPRB2005}:

\begin{equation}
\begin{aligned}
C_{11} + C_{12} = & \frac{1}{\sqrt{3} c_0} \frac{\partial^2 E}{\partial a^2} \\
C_{33}          = & \frac{2c_0}{\sqrt{3} a_0^2} \frac{\partial^2 E}{\partial c^2} \\
C_{13}          = & \frac{1}{\sqrt{3} a_0} \frac{\partial^2 E}{\partial a \partial c} \\
\end{aligned}
\label{elastic_constants_graphite_eq}
\end{equation}

\noindent
where the derivatives are evaluated at equilibrium. In order to calculate these derivatives, we compute the total energy for several values of $a$ and $c$ around the equilibrium values $a_0$ and $c_0$. The resulting energy landscape is then fitted to two-dimensional fourth order Taylor expansion and the second order coeficients provide the required derivatives. Exfoliation energies per area for graphite and NLG are calculated using the following equation

\begin{equation}
\Delta E_{ex}(N) = -\frac{(E_{N} - E_{N-1} - E_{graphene})}{A}
\end{equation}

\noindent and binding energies per area per layer are calculated using the following equation

\begin{equation}
\Delta E_{b}(N) = -\frac{(E_{N}/N -  E_{graphene})}{A}
\end{equation}

\noindent
where $E_{N}$ is the total energy per unit cell of NLG or graphite, and $E_{graphene}$ is the graphene total energy per unit cell, and $A$ is the basal area of the unit cell. All energies are calculated for fully relaxed systems, therefore the calculated exfoliation energies for NLG include surface relaxation effects. In the case of graphite, in the limit $N \to \infty$, the exfoliation energy coincides with the binding energy \cite{Bjorkman2012}.

\section{\label{sec:level1}Results and Discussion}

\subsection{Graphite}

Results for lattice constants, elastic constants, binding energies and vibrational frequencies of graphite are shown in Table \ref{tabela_grafite} for different functionals. Note that LDA underestimates the value of $a_0$ and GGA overestimates the value of $c_0$, as expected. All vdW functionals give good values of $a_0$, with the vdw-DF2 showing the largest deviation from the experimental value. For $c_0$, most vdW functionals show good agreement with experiment. Exceptions are the vdW-DF1 and vdW-DF2 functionals, which show the largest deviations.

All calculated $C_{13}$ values are negative, in agreement with previous theoretical works \cite{MounetPRB2005}. Experimental values lie in the range 0-15 GPa, with an error bar of 3 GPa, so a negative value cannot be entirely discarded. Note that most vdW functionals give elastic constants with absolute values larger than LDA as these elastic constants are related to the interactions between graphene layers. Finally, GGA, vdW-DF1 and vdW-DF2 provide inaccurate descriptions of $C_{11} + C_{12}$ because these functionals do not describe very well the interplanar distance. The use of experimental values $a_0$ and $c_0$ in Eq. (\ref{elastic_constants_graphite_eq}) may improve the evaluation of these elastic constants, as discussed in Ref. \onlinecite{MounetPRB2005}.

\begin{table*}
\centering
\caption{Calculated lattice parameters $a_0$ and $c_0$, elastic constants, $C_{13}$, $C_{33}$ and $C_{11}+C_{12}$, binding energies $\Delta E_b$, frequencies of shear and layer breathing modes (LBM) of graphite and mean relative dispersion $\Delta$ and $\Delta_{freq}$ (explained in text), as given by different functionals employed on this work.}
\vspace{0.5cm}
\begin{ruledtabular}
\begin{tabular*}{\textwidth}{r|ccccccccccc}

\footnotetext[1]{Mean of experimental values. These values were used as reference values to evaluate the performance of the vdW functionals.}
\footnotetext[2]{Ref. \onlinecite{Hanfland1989PRB}}
\footnotetext[3]{Ref. \onlinecite{PierreBook2000}}
\footnotetext[4]{Ref. \onlinecite{BosakPRB2007}}
\footnotetext[5]{Ref. \onlinecite{EklundCarbon1995}}
\footnotetext[6]{Ref. \onlinecite{TanNatMat2012}}
\footnotetext[7]{Ref. \onlinecite{Zacharia2004OPRB}}
\footnotetext[8]{Ref. \onlinecite{Liu2012PRB}}
\footnotetext[9]{Ref. \onlinecite{BenedictChemPhysLett1998}}
\footnotetext[10]{Ref. \onlinecite{Wang2015NatComm}}
\footnotetext[11]{Ref. \onlinecite{Boschetto2013NanoLet}}

 & $a_0$ & $c_0$ & $C_{13}$ & $C_{33}$ & $C_{11}+C_{12}$ & $\Delta E_{b}$ & Shear Mode & LBM & $\Delta$ & $\Delta_{freq}$ \\ % Note a separação de col. e a quebra de linhas
 &   (\AA) & (\AA)  &   (GPa)  &  (GPa)   &      (GPa) & (J/m$^{2}$) & (cm$^{-1}$) & (cm$^{-1}$)  & (\%) & (\%)\\
\hline                               % para uma linha horizontal
exp. values  & $2.459$\footnotemark[2] & $6.706$\footnotemark[2] & $15$\footnotemark[3] & $36.5$\footnotemark[3] &  $1240$\footnotemark[3] & 0.32$\pm$0.03\footnotemark[7], 0.19$\pm$0.01\footnotemark[8]  & 42\footnotemark[5] & 127\footnotemark[5] & - & - \\
&  &  & $0$\footnotemark[4]  & $38.7$\footnotemark[4] & $1248$\footnotemark[4] & 0.21$\pm$0.06\footnotemark[9], 0.33\footnotemark[10] & 43.5\footnotemark[6], 44.03\footnotemark[11] & & & \\

Reference values\footnotemark[1] & - & 6.706 & - & 37.6 & - & 0.2625 & 43.2 & 127 & - & - \\

This work & & & & & & & & & & \\

GGA          & 2.46  & 8.27 & -0.5 & 3.7 & 1005.9  & 0.01 & 10.4  & 36.7 & 71.4 & 73.5 \\
LDA          & 2.45  & 6.62 & -2.8 & 32.0 & 1309.5 & 0.15 & 44.2 & 118.8 & 13.6 & 4.4 \\
RVV10        & 2.48  & 6.71  & -4.3 & 44.7 & 1210.3 & 0.43 &  42.4 & 145.0 & 19.7 & 8.0 \\
vdW-DF1       & 2.47  & 7.15 & -2.8 & 27.1 & 1124.9 & 0.33 &  28.2 & 108.7 & 21.9 & 24.5 \\
vdW-DF1-c09   & 2.46  & 6.45  & -4.0 & 48.2 & 1299.4 & 0.47 &  50.9 & 151.5 & 29.7 & 18.6 \\
vdw-DF1-cx    & 2.46  & 6.55 & -3.8 & 47.7 & 1276.9 & 0.40 &  47.1 & 137.5 & 19.8 & 8.7 \\
vdW-DF1-optB86b  & 2.46  & 6.61 & -3.9 & 48.1 & 1255.3 & 0.44 &  44.8 & 138.9 & 22.0 & 6.6 \\
vdW-DF1-optB88  & 2.47  & 6.68 & -3.7 & 42.2 & 1230.2 & 0.43 &  42.2 & 138.6 & 17.6 & 5.7 \\
vdW-DF2      & 2.49  & 7.05 &  -3.3 & 34.7 & 1111.1 & 0.33 &  31.9 & 123.4 & 13.5 & 14.5 \\
vdW-DF2-B86r & 2.47  & 6.63 & -4.0 & 40.7 & 1250.1 & 0.37 &  44.6 & 137.6 & 12.4 & 5.8 \\
vdW-DF2-c09  & 2.46  & 6.53 & -4.1 & 42.0 & 1277.6 & 0.34 &  54.3 & 141.9 & 16.3 & 18.8 \\

Other theoretical works & & & & & & & & & & \\

GGA \cite{MounetPRB2005} & 2.46  & 8.49 & -0.46 & 2.4 & 976  & - & 8  & 28 & - & - \\
LDA \cite{MounetPRB2005} & 2.44  & 6.68 & -2.8 & 29 & 1283  & - & 44  & 113 & - & - \\
PBE+D2 \cite{Rego2015PhysConMat} & 2.461  & 6.444 & - & 22.04 &  - & 0.34 & -  & - & - & - \\
PBE+D3 \cite{Rego2015PhysConMat} & 2.464  & 6.965 & - & 13.09 &  - & 0.29 & -  & - & - & - \\
PBE+D3-BJ \cite{Rego2015PhysConMat} & 2.464  & 6.745 & - & 17.16 &  - & 0.32 & -  & - & - & - \\
PBE+TS \cite{Rego2015PhysConMat} & 2.458  & 6.665 & - & 34.61 &  - & 0.50 & -  & - & - & - \\
PBE+TS+SCS \cite{Rego2015PhysConMat} & 2.461  & 6.633 & - & 26.79 &  - & 0.33 & -  & - & - & - \\
ACFDT-RPA \cite{Leb2010PhysRevLet} & -  & 6.68 & - & 36 &  - & 0.29 & -  & - & - & - \\
QMC \cite{SpanuPRL2009} & -  & 6.852 & - & - &  - & 0.34$\pm$0.03 & -  & - & - & - \\
LDA \cite{ChenScienRep2013} & -  & 6.668 & - & 23.67 &  - & 0.14 & -  & - & - & - \\
LDA+D2 \cite{ChenScienRep2013} & -  & 5.978 & - & 95.23 &  - & 0.70 & -  & - & - & - \\
PBE \cite{ChenScienRep2013} & -  & 8.838 & - & 1.390 &  - & 0.005 & -  & - & - & - \\
PBE+D2 \cite{ChenScienRep2013} & -  & 6.462 & - & 42.44 &  - & 0.37 & -  & - & - & - \\
optPBE+D2 \cite{ChenScienRep2013} & -  & 6.894 & - & 33.15 &  - & 0.39 & -  & - & - & - \\
vdW-DF1-optB88 \cite{ChenScienRep2013} & -  & 6.712 & - & 40.37 &  - & 0.42 & -  & - & - & - \\
rPW86/vdW-DF2 \cite{ChenScienRep2013} & -  & 7.048 & - & 34.11 &  - & 0.32 & -  & - & - & - \\

\end{tabular*}
\end{ruledtabular}
\label{tabela_grafite}
\end{table*}

\begin{figure}
\includegraphics[width=1.0 \linewidth]{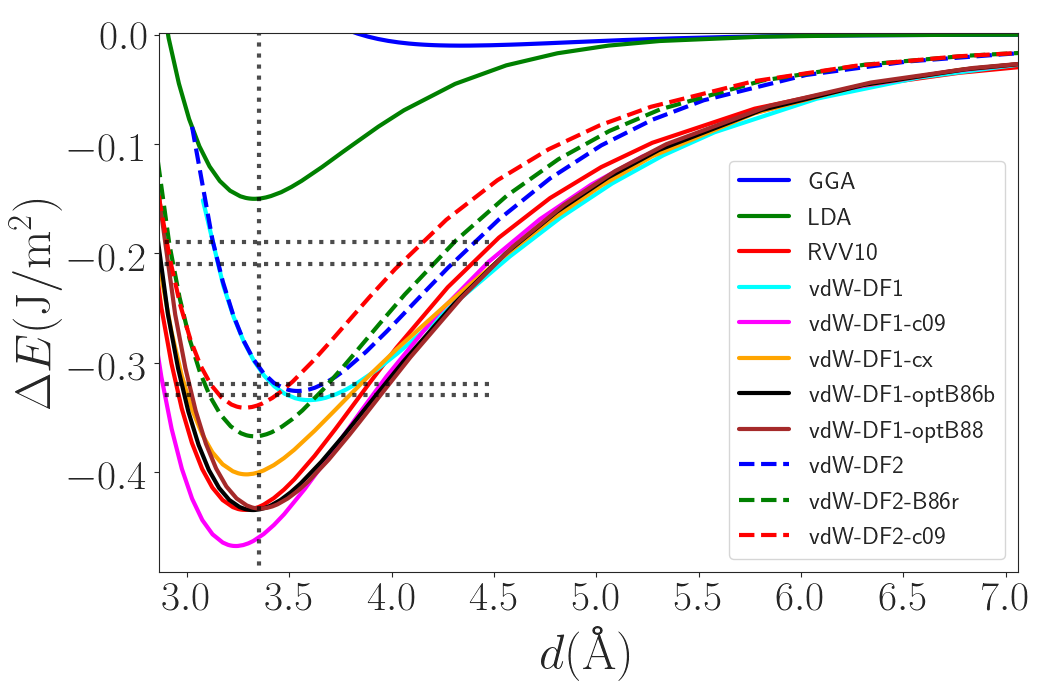}
\caption{Total energy difference per unit area between graphite and graphene as a function of the interlayer distance for different functionals. The vertical dotted line corresponds to the experimental interlayer distance and horizontal dotted lines correspond to the experimental values reported in Table \ref{tabela_grafite}. The zero energy corresponds to graphene energy per unit area.}
\label{E_vs_c}
\end{figure}

In Fig. \ref{E_vs_c}, the total energy difference (per area and per layer) between graphite and graphene as a function of interlayer separation is shown. The minimum value is, by definition, the negative of the binding energy, whose values for different functionals are also reported in Table \ref{tabela_grafite}. Experimental values for the exfoliation (or binding) energies are displayed by horizontal dotted lines. We can see that the LDA and GGA functionals give very small values for this quantity in comparison with experimental results while all vdW functionals tend to overestimate it. Our results show a good agreement with other theoretical calculations (reported in Table \ref{tabela_grafite}) using Quantum Monte Carlo \cite{SpanuPRL2009}, adiabatic-connection fluctuation-dissipation theorem in the random phase approximation (ACFDT-RPA) \cite{Leb2010PhysRevLet}, DFT-D and TS-vdW methods \cite{ChenScienRep2013, Rego2015PhysConMat} and vdW functionals \cite{ChenScienRep2013}. For vdW functionals, it is possible to see two families of curves in the range from $4\mathrm{\AA}$  to $7\mathrm{\AA}$. The solid lines are the vdW-DF type functionals and dashed lines are vdW-DF2 type functionals. Each set converges to a particular asymptotic curve, due to the different asymptotic behaviors of DF1 and DF2 kernels in Eq. \ref{Non_local_corr_energy} \cite{lee2010PRB}.

By using the frozen-phonon technique, the frequencies of the layer shear and breathing modes (LBM) are calculated and reported in Table \ref{tabela_grafite}. In this technique, phonon frequencies are calculated from the curvature of the energy versus displacement curves. The atomic displacements are performed according to a given normal mode, such that the energy differences with respect to the equilibrium configuration are solely due to this mode \cite{KuncPRL81}. Several vdW functionals (and, interestingly, also LDA) display frequencies with reasonably good agreement with experiment \cite{EklundCarbon1995, TanNatMat2012}.

To evaluate the performance of different functionals we calculated the mean relative standard deviation $\Delta = (\sum_i \sqrt{(x_{calc, i}-x_{ref, i})^2}/x_{ref, i})/M$ ($M$ is the number of evaluated properties) between calculated and reference experimental values ($x_{ref,i}$) for each functional using the following quantities: $c_0$, $C_{33}$, $\Delta E_{b}$ and the frequencies of shear and layer breathing modes, as these quantities strongly depend on vdW interactions. $C_{13}$ was not taken into account as all calculated values are far from experimental values reported in Table \ref{tabela_grafite}. The reference experimental values are taken as the mean of avaiable experimental data and they are also shown in Table \ref{tabela_grafite}. Based on this analysis, the functional with best overall performance would be vdW-DF2-B86r. However, since our main objective in this work is the calculation of vibrational frequencies in NLG, we also calculate the mean deviation restricted to the LBM and shear-mode vibrational frequencies, labeled $\Delta_{freq}$ in Table \ref{tabela_grafite}. Based on this more restricted analysis the vdW-DF1-optB88 functional shows the best performance with a standard deviation of 5.7$\%$. It should also be mentioned that, even though the LDA functional also has a good performance, it gives a quite poor description of the binding energy of graphite, so we choose the vdW-DF1-optB88 for the study of the vibrational properties of NLG.

\begin{figure}
\includegraphics[width=1.0 \linewidth]{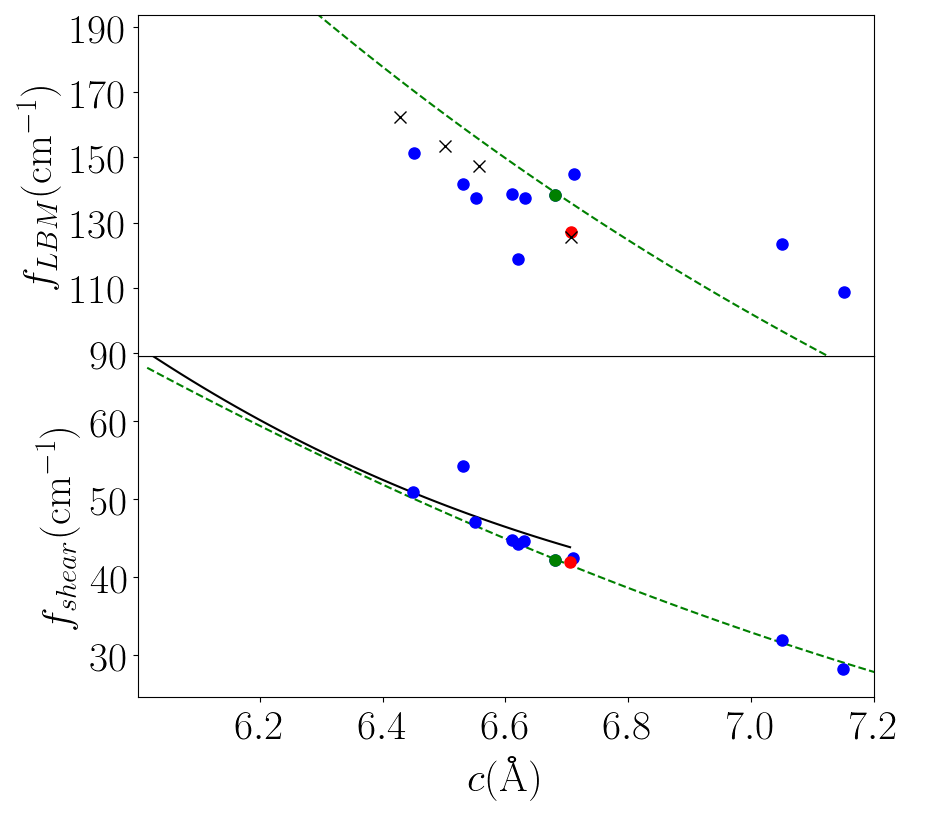}
\caption{Calculated frequencies for the LBM and shear mode of graphite as a function of the equilibrium out-of-plane lattice constant for each functional used in this work. The data points are taken from Table \ref{tabela_grafite} and are represented by green and blue points. The green points are results given by the vdW-DF1-optB88 functional, while blue points are results from other functionals (GGA not included). The green dashed lines correspond to frequencies calculated using the vdW-DF1-optB88 functional for different interlayer distances, which do not correspond to the equilibrium value for this functional. Red points are experimental values from Ref.\onlinecite{EklundCarbon1995}. The black line and black crosses are experimental results from Refs. \onlinecite{Hanfland1989PRB} and \onlinecite{Alzyab1988PRB}, respectively.}
\label{Freq_vs_dist_interplanar}
\end{figure}

By analizing the results in Table \ref{tabela_grafite}, we notice a correlation, shown in Fig. \ref{Freq_vs_dist_interplanar}, between equilibrium $c$ lattice parameters and calculated frequencies using different functionals (green and blue circles). Frequencies tend to be higher for smaller interlayer distances. As a matter of fact, this seems to be a fairly universal trend, as it is approximately followed by experimental results in graphite at zero (red circles) and non-zero pressures (black line and crosses), and also by calculations using the vdW-DF1-optB88 functional and varying the interplanar distance (green dashed line). As one can see, the results for this functional follow very well the experimental trends as a function of pressure.

\subsection{N-Layer Graphene}

\begin{figure}
\includegraphics[width=1.0 \linewidth]{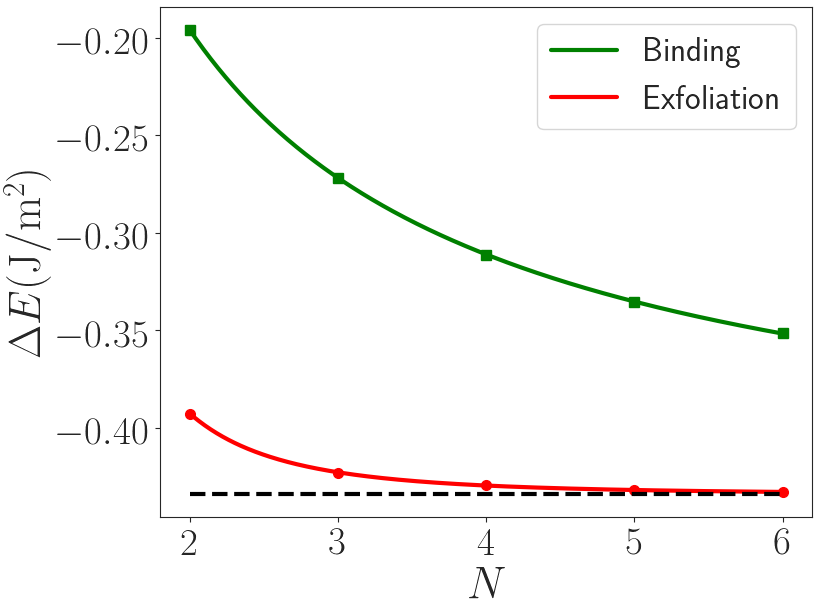}
\caption{Binding (green square dots) and exfoliation (red circular dots) energies per unit area of NLG as a function of the number of layers using vdW-DF1-optB88 functional. Dashed line corresponds to graphite binding energy and solid lines are fits given by Eq. \ref{fit_exf_energy}.}
\label{E_NLG}
\end{figure}

By using the vdW-DF1-optB88 functional, we evaluated the structural and vibrational properties of multilayer graphene, from the bilayer ($N = 2$) up to 6 stacked layers. In all cases, the layers are AB (Bernal) stacked. The calculated binding and exfoliation energies are shown in Fig. \ref{E_NLG}. As expected, they both converge to the graphite value with increasing number of layers. The green and red curves are fits given by the following equation:

\begin{equation}
\Delta E_{ex(b)}(N) = \frac{A}{N^b} + \Delta E_{ex(b)}(\infty)
\label{fit_exf_energy}
\end{equation}

\noindent where $\Delta E_{ex(b)}(\infty)$ corresponds to graphite. For NLG, by using the vdW-DF1-optB88 functional, we find $b = 0.90$ for the binding case and $b = 3.20$ for the exfoliation case. Therefore, the binding energy varies more slowly with $N$, but both binding and exfoliation energies yield the graphite exfoliation energy $\Delta E_{ex}(\infty)$ in the limit $N \to \infty$, as expected. In fact, the values of $\Delta E_{ex}(\infty)$ as given by each fit differ from the actual value by less than $0.014 J/m^2$. We have also calculated the interlayer distances in all cases and variations with respect to graphite values are smaller than $0.015 \mathrm{\AA}$.

We also evaluated the interlayer distance and the energy difference between AA and AB stackings for 2LG and graphite (Table \ref{table_2LG_AA_AB}) due to the increasing importance of studies in twisted bilayer graphene (t2LG) \cite{cao2018Nature}. As t2LG is composed by AA and AB domains the interlayer distance an coupling strength between the layers varies through this material \cite{Wu2014NatComm}.

\begin{table*}
\caption{Interlayer distance and energy difference between AA and AB stackings ($E_{AA} - E_{AB}$) for graphite and 2LG.}
\vspace{0.5cm}
\begin{ruledtabular}
\begin{tabular*}{\textwidth}{rr|ccc}

 & & Interlayer Distance ($\mathrm{\AA}$) & & $E_{AA} - E_{AB}$ (meV/atom) \\
 & & AA & AB & \\
\hline	
Graphite     & vdW-DF1-optB88              & 3.528 & 3.343 & 10.5 \\
             & vdW-DF1-optB86b             & 3.522 & 3.316 & 11.3 \\
\hline	             
2LG          & vdW-DF1-optB88              & 3.462 & 3.362 & 4.7 \\
             & vdW-DF1-optB86b             & 3.522 & 3.316 & 5.0 \\
             & Exp \cite{brown2012NanoLet} &   -   & 3.35  & -   \\

\end{tabular*}
\end{ruledtabular}
\label{table_2LG_AA_AB}
\end{table*}

\begin{figure}
\includegraphics[width=1.0 \linewidth]{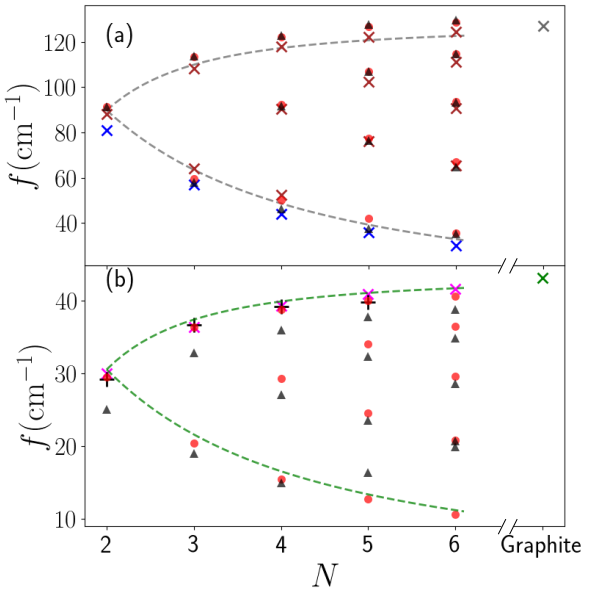}
\caption{Layer breathing (a) and shear (b) mode frequencies of NLG and graphite at the $\Gamma$ point. Red dots are calculations using the vdW-DF1-optB88 functional and black triangles are calculations including first order perturbation theory (see text). Experimental values given in Ref. \onlinecite{LuiNanoLet2014} (blue crosses), Ref. \onlinecite{TanNatMat2012}(pink crosses), Ref. \onlinecite{Boschetto2013NanoLet}(black plus crosses), Ref. \onlinecite{EklundCarbon1995} (gray crosses) and Ref. \onlinecite{Lui2013PRB} (brown crosses). Dashed lines are predictions using the experimental values of graphite and Eq. (\ref{Freq_NLG_grafite}) for the upper and lower branches.}
\label{NLG_freqs}
\end{figure}

Phonon frequencies are evaluated by using the method of finite differences. Starting from the relaxed structure in each case, displacements of the $i$-th sheet and in the $\mu$ direction ($\mu=x$, $y$ or $z$) are applied and the forces on the $j$-th sheet and $\nu$ direction ($\nu=x$, $y$ or $z$) are computed. In this work, the evaluated optical modes correspond to rigid displacements of the graphene sheets without any internal displacements.  Therefore, the corresponding frequencies are evaluated at the $\Gamma$ point of the Brillouin Zone, and they are shown in Fig. \ref{NLG_freqs}. Red dots are calculations using vdW-DF1-optb88 functional within the harmonic approximation. Anharmonicity effects are also taken into account by using first-order perturbation theory (see below) and the corresponding results are represented by black triangles. The remaining dots are experimental results (details are given in the figure caption). Wherever experimental results are avaiable, the agreement between theory and experiment is quite good. Therefore, it is safe to assume that our theoretical results have good predictive power for the remaining cases.

\begin{figure}
\includegraphics[width=1.0 \linewidth]{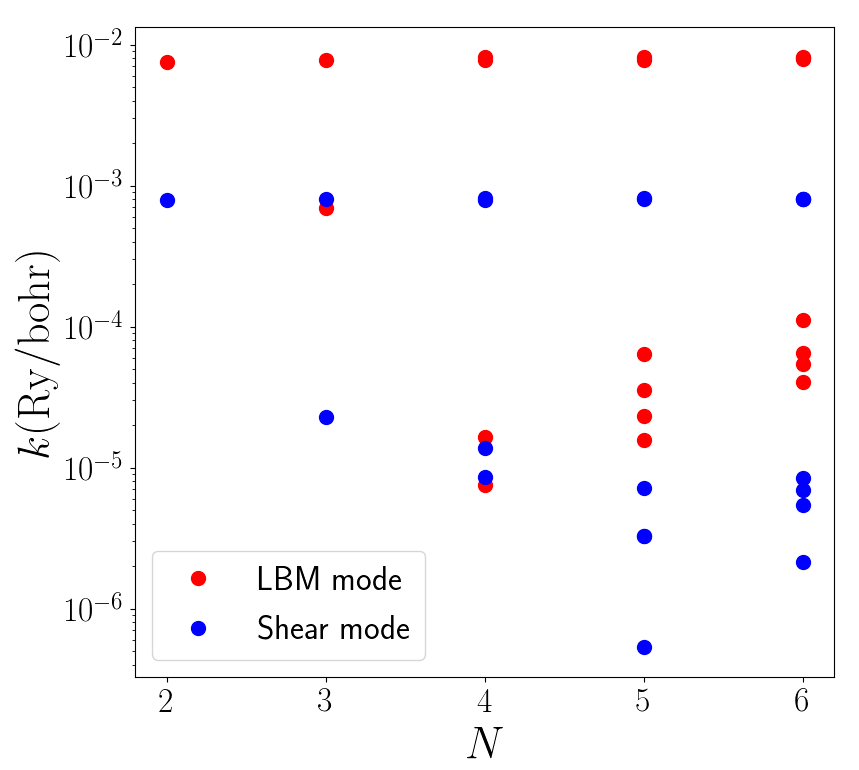}
\caption{Calculated force constants for NLG given by vdW-DF1-optB88 functional. Red circles correspond to the LBM and blue circles to shear modes.}
\label{NLG_force_constants}
\end{figure}

The dashed lines in Fig. \ref{NLG_freqs} are obtained by using a model of interactions of equal magnitude between nearest-neighbor layers in which our system is  modeled as $N$ masses connected by $N-1$ springs to compute the frequencies of the LBM and shear modes. In this model, the frequencies are given by \cite{TanNatMat2012, Boschetto2013NanoLet}

\begin{equation}
\omega_{N, i}^2 = \frac{\omega^2_{graphite}}{2} \left[ 1 - \mathrm{cos} \left( \frac{\pi(i-1)}{N} \right) \right],
\label{Freq_NLG_grafite}
\end{equation}

\noindent where $i=2,...,N$ is the mode index, $N$ is the number of layers and $\omega_{graphite}$ is the frequency of the corresponding mode in graphite. This equation can be used to predict the frequencies of NLG from the corresponding graphite frequencies, as shown by the dashed lines in Fig. \ref{NLG_freqs}. By using the average of available experimental data for graphite frequencies, we have plotted, for clarity, only the curves corresponding to the upper and lower branches of Eq. \ref{Freq_NLG_grafite} ($i = N$ and $i = 2$, respectively) \cite{TanNatMat2012, LuiNanoLet2014, Boschetto2013NanoLet}. For both modes, this simple analytical model shows good agreement with our calculated results and available experimental data. This suggest that it is a good approximation to consider only interactions between nearest-neighbor layers. Indeed, as a by-product of our finite-difference phonon calculations, we obtain the effective force constants between pairs of layers for both normal (LBM) and tangential (shear) displacements. They are shown in Fig. \ref{NLG_force_constants} and one can notice that, for all systems, the effective force constants between nearest-neighbor layers is typically one order of magnitude larger than for second neighbors and beyond.

Finally, we also study the effects of anharmonicity by applying first-order perturbation theory to the finite differences results. The total energy vs. displacement curve in the harmonic approximation is written as an oscillator potential energy:

\begin{equation}
V_0 = \frac{1}{2m}\omega_{0, shear}^2(x^2 + y^2) + \frac{1}{2m}\omega_{0, LBM}^2z^2,
\end{equation}

\noindent which is the potential energy of a 3D anisotropic quantum harmonic oscilator whose eigenvectors are $\Ket{n_x, n_y, n_z}$ and eigenvalues are $E^{(0)}(n_x, n_y, n_z) = \hbar \omega_{0, shear}(n_x + n_y + 1) + \hbar \omega_{0, LBM} (n_z + 1/2)$. The perturbation potential is given by

\begin{equation}
V'= Ax^4 + By^4 + Cz^4.
\end{equation}

\noindent Note that terms of third order on the position operators give a zero contribution to the first order correction. The corrections to the frequencies are $\Delta \omega = \Delta E^{(1)}/\hbar$, where $E^{(1)}(n_x, n_y, n_z)=\Braket{n_x, n_y, n_z|V'|n_x, n_y, n_z}$ are the first order corrections given by

\begin{equation}
\begin{aligned}
& E^{(1)}(n_x, n_y, n_z)  = \\ 
& \left( \frac{\hbar}{2m\omega_{0,shear}} \right)^2 \left( 3A \left( n_x^2 + n_x + \frac{1}{2} \right) + 3B \left( n_y^2 + n_y + \frac{1}{2} \right) \right) \\
& + \left( \frac{\hbar}{2m\omega_{0,LBM}} \right)^2  6C \left( n_z^2 + n_z + \frac{1}{2} \right).
\end{aligned}
\end{equation}

\noindent For the LBM, the corrections are about a few cm$^{-1}$ and, for the lowest branch in Fig. \ref{NLG_freqs} they yield a better agreement with data from Ref. \onlinecite{LuiNanoLet2014}. For the shear mode, the corrections are larger than for the LBM, but they lead to a worse agreement with experimental results.

\section{\label{sec:level1}Conclusions}

We have evaluated the performance of different vdW functionals on the description of structural, mechanical and vibrational properties of graphite. Our results indicate that, although the vdW-DF2B86r functional has the best overall performance, the vdW-DF1-optB88 is more adequate for vibrational properties. Many other functionals, such as vdW-DF1-opt86b, vdW-DF-cx and RVV10, show very good agreement with experimental results as well. Even LDA performs well for a few structural properties, but it provides too small exfoliation energies.

To study the vibrational properties of NLG, we used the vdW-DF1-optB88 functional. Calculated vibrational frequencies match with great accuracy available experimental results and predictions of analytical models based on nearest-neighbor layer interactions. This suggests that our methodology has predictive power for the frequencies which have not been measured so far. Our results reinforce the adequacy of recently developed vdW functionals and their importance to the study of graphene-based layered systems, other layered two-dimensional materials and van-der-Waals heterostructures and illustrate the need of inclusion of vdW interactions in order to correctly describe these properties, especially those related to out-of-plane atomic displacements.

\section*{Acknowledgements}

We acknowledge the financial support from the Brazilian agencies CNPq, CAPES, FAPERJ and INCT-Nanomateriais de Carbono. We also thank DIMAT-Inmetro and COPPE-UFRJ for the computational resources employed on this work.

%----------------------------------------------------------------------------------------
%	REFERENCES
%----------------------------------------------------------------------------------------

\bibliography{references}{}
\bibliographystyle{unsrt}

\end{document}